\documentclass{he_symp}
\usepackage{psfig}
\newcommand{\gapr}{\raisebox{-.6ex}{\mbox{
$\stackrel{>}{\mbox{\scriptsize$\sim$}}\:$}}}

\newcommand{\vv}[1]{\mbox{\boldmath $#1$}}
\def\tef{T_{\rm eff}}

\begin{document}
\title{Modeling Neutron Star Atmospheres}
\author{V.E. Zavlin\inst{1} \and G.G. Pavlov\inst{2}}  
\institute{Max--Planck--Institut f\"ur extraterrestrische Physik,
 Giessenbachstra{\ss}e, 85748 Garching, Germany
\and  The Pennsylvania State University, 525 Davey Lab,
University Park, PA 16802, USA}
\maketitle

\begin{abstract}
Models of thermal emission of neutron stars, presumably formed in their
atmospheres,
are needed to infer
the surface temperatures, magnetic fields, chemical composition,
and neutron star masses and radii from the observational data.
This information, supplemented
with  model equations of state
and neutron star cooling models,
is expected to move us further in understanding the fundamental 
properties of the
superdense matter in the neutron star interiors.
The neutron star atmospheres are very different from those of
usual stars due to the immense gravity and huge magnetic
fields. 
In this presentation
we review the current status of the neutron star 
atmosphere modeling and
present most important 
results.
\end{abstract}

\section{Introduction}
A systematic study of
X-ray emission from isolated neutron stars (NSs),
including radio pulsars,
has started
after the launch of the {\sl Einstein} and {\sl EXOSAT} space
observatories.
These studies have shown that, generally, there are two different
components of the NS X-ray emission --- thermal and nonthermal.
The nonthermal component with a power-law spectrum,
observed from many radio pulsars,
is believed to originate from the pulsar's magnetosphere,
while the thermal component is emitted from
the NS surface layers (atmospheres).
The thermal radiation is particularly interesting
because it can provide 
important 
information about the NS: its surface temperature,
magnetic field,
and chemical composition, as well as the NS radius 
and mass.
Measuring these parameters for a sample of NSs 
is necessary for studying the
thermal evolution of NSs and constraining the
equation of state and composition
of the superdense matter 
in the NS interiors (see the review by Yakovlev et al. in
these Proceedings).

Thermal X-ray emission from the
NS surface 
had been discussed
by Chiu \& Salpeter and Tsuruta in 1964, 
before NSs were discovered, and well before
their thermal emission was 
detected
with {\sl Einstein} and {\sl EXOSAT} 
(e.g., Cheng \& Helfand 1983; Brinkmann \& \"Ogelman 1987;
C\'ordova et al.~1989;
Kellet at al.~1987). 
Many new observational results on the NS thermal radiation
were obtained in 1990' with the {\sl ROSAT}, {\sl ASCA},
and {\sl EUVE} satellites
(see Becker \& Pavlov 2002 for a review).
Thermal radiation from a few NSs
 was also detected in the optical-UV energy range
with the {\sl Hubble Space Telescope} (e.~g., Pavlov, Stringfellow \&
C\'ordova 1996; Walter \& Matthews 1997).
Currently operating X-ray observatories,
{\sl Chandra} and {\sl XMM}-Newton,
are providing new excellent data on X-rays from NSs 
(see the contributions 
by Weisskopf, Becker, and Pavlov, Zavlin \& Sanwal
in these Proceedings).

To interpret these observations, one needs reliable
models for the NS thermal radiation.
The fact that the spectrum of radiation emergent from a NS
atmosphere can be very different from a blackbody spectrum,
and the angular distribution can be very far from isotropic,
particular in a strong magnetic field, has been recognized
long ago. For instance,
Pavlov \& Shibanov (1978) calculated spectra and angular distributions
of radiation 
from a strongly magnetized NS atmosphere
assuming the source function
grows inward linearly
in the emitting layers. First self-consistent
models for NS atmospheres 
were developed
by Romani (1987), for low magnetic fields,
and Shibanov et al.\ (1992), for strong magnetic fields.
Since then, various aspects of the NS atmosphere modeling
have been investigated in many papers.     Below we will overview
the current status 
of this field and 
summarize some important results.

\section{ Distinctive features of NS atmospheres}
The NS thermal radiation emerges,
as in usual
stars, in superficial layers which can be in a gaseous state
(atmosphere) or a condensed state (liquid or solid
surface),
depending on surface temperature, magnetic field and chemical
composition. 
A condensed surface forms at low temperatures and very strong
magnetic fields.
For instance,
according to the estimates by Lai \& Salpeter (1997), 
hydrogen is condensed in surface layers if
$T\la 1\times 10^5$ K at $B=1\times 10^{13}$ G  
($T\la 1\times 10^6$ K at
$B=5\times 10^{14}$ G).
At higher temperatures and/or lower magnetic fields,
hydrogen does not condensate
and forms an atmosphere.

In both usual stars and NSs the properties of radiation
emergent from an atmosphere strongly depend on its
chemical composition. The difference between usual stars and
NSs is that in NSs we expect the emitting layers to be comprised
of just one, lightest available, chemical element, not a mixture of
elements, because heavier elements sink into deeper layers due to
the immense NS gravitation (Alcock \& Illarionov 1980).
For instance, even a small amount of hydrogen (with a surface density
of $\sim 
1$ g cm$^{-2}$) is sufficient for the 
radiation to be indistinguishable from that
emitted from a purely hydrogen atmosphere.
Such an amount of hydrogen, $\gapr 10^{-20}M_\odot$, can 
be delivered onto the NS surface 
by, e.~g., accretion from the interstellar medium and/or
fallback of a fraction of the envelope ejected
during the supernova explosion. 
If no hydrogen is present at
the surface, 
a heavier chemical element 
is responsible for
the radiative properties of the NS atmosphere.
However, a mixture of elements can be observed in the emitting layers
if the NS is experiencing accretion with such a rate that
the accreting matter is supplied faster than the gravitational
separation occurs.
Therefore, it is important to construct the NS atmosphere
models for a 
variety of surface compositions, both `pure' and mixed.

\begin{figure}
\centerline{\psfig{file=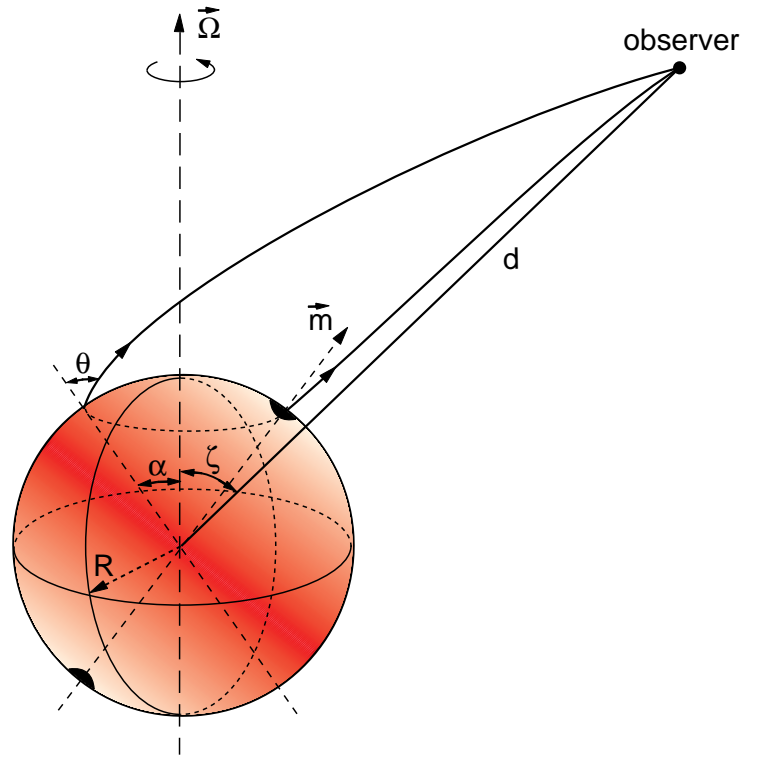,height=12cm,clip=} }
\caption{
Bending of photon trajectories in a strong
gravitational field near the surface of a NS with rotational
and magnetic axes $\vv{\Omega}$ and $\vv{m}$, respectively. 
}
\label{sketch}
\end{figure}

The gravitational separation of elements
is one of the consequences of the enormous gravity at the NS surface,
with the gravitational acceleration $g\sim 10^{14}-10^{15}$ cm s$^{-2}$.
The gravity makes NS atmospheres
very thin, 
$\sim 0.1-10$ cm,
and dense, $\rho\sim 10^{-2}-10^2$ g cm$^{-3}$.
Such a high
density causes
strong nonideality effects (pressure ionization,
smoothed spectral features) which must be taken into account
(e.~g., Pavlov et al.\ 1995).
In addition, the strong gravitational field
bends the photon trajectories near the NS surface
(Pechenick, Ftaclas \& Cohen 1983),
as illustrated by the sketch in Figure~\ref{sketch}. 
This effect depends on the gravitational parameter 
$g_{\rm r}=(1-2GM/c^2R)^{1/2}$, and it can even
make the whole NS surface visible if the NS is massive enough,
$M > 1.8\, (10\, {\rm km}/R)\, M_\odot$. In particular,
the gravitational
bending 
strongly affects the observed pulsations
of thermal emission
(Zavlin,  Shibanov \& Pavlov 1995a).

Huge magnetic fields, $B\sim 10^{11}-10^{14}$ G, in the surface
layers of many NSs change
the properties of the atmospheric
matter and the emergent radiation even more drastically. 
Strongly magnetized NS atmospheres are essentially anisotropic,
with radiative opacities depending on the magnetic field and
the direction and polarization
of radiation. Moreover, since the ratio of the cyclotron
energy, $E_{\rm ce}=\hbar eB/m_{\rm e}c$, to the Coulomb energy
is very large (e.g.,
$\gamma\equiv E_{\rm ce}/(1\, {\rm Ry}) = 850\, (B/10^{12}\, {\rm G})$
for a hydrogen atom),
the structure of atoms is strongly distorted
by the magnetic field. For instance, the binding (ionization) energies
of atoms are increased by a factor of $\sim \ln^2\gamma$ (e.~g.,
the ionization potential of a hydrogen atom is 
about 
310 eV at $B=10^{13}$ G). 
This, in turn, significantly modifies ionization
equilibrium of the NS atmospheric matter.
Another important effect is that 
the heat conductivity of the NS crust is anisotropic (it is higher
along the magnetic field). This results in a nonuniform surface
temperature distribution (Greenstein \& Hartke 1983), which
leads to pulsations of the thermal radiation
due to NS rotation.

\section{Atmosphere models with low magnetic fields}
Some NSs (e.g., millisecond pulsars) have relatively low magnetic fields,
$B\la 10^9$ G. 
In such magnetic fields the electron
cyclotron energy,
$E_{\rm ce}\la 0.01$ keV, is lower than the binding energy
of atoms and thermal energy of particles.
As a result, the effect
of the field on the radiative opacities 
and
emitted spectra is negligible,
at X-ray energies, $E\ga 0.1$ keV.
Hence, to model low-field NS atmospheres,
one can merely put $B=0$.
These models are applicable to millisecond pulsars, 
NS transients in quiescence
(e.g., Rutledge et al.\ 2002), and, perhaps, to some
radio-quiet isolated NSs.
One should remember,
however, that 
the magnetic field effects on spectral opacity at optical wavelengths
must be included even for these low fields if the models
are applied for interpreting the optical spectra.

First models of low-field NS atmospheres
were presented in the pioneering work by Romani (1987). Since then,
models for various surface chemical compositions have been
developed by Rajagopal \& Romani (1996), Zavlin, Pavlov
\& Shibanov (1996), Pavlov \& Zavlin (2000a),
Werner \& Deetjen (2000),
Pons et al.\ (2002), G\"ansicke, Braje \& Romani (2002). 

\subsection{General approach}
As the
thickness of a NS atmosphere is much smaller
than 
the NS radius, $R\approx 10$ km, NS atmospheres 
can be considered in the plane-parallel approximation.
In addition, rather high 
densities of the surface layers allow one to consider the
atmospheres to be in local thermodynamic equilibrium.

\begin{figure}
\centerline{\psfig{file=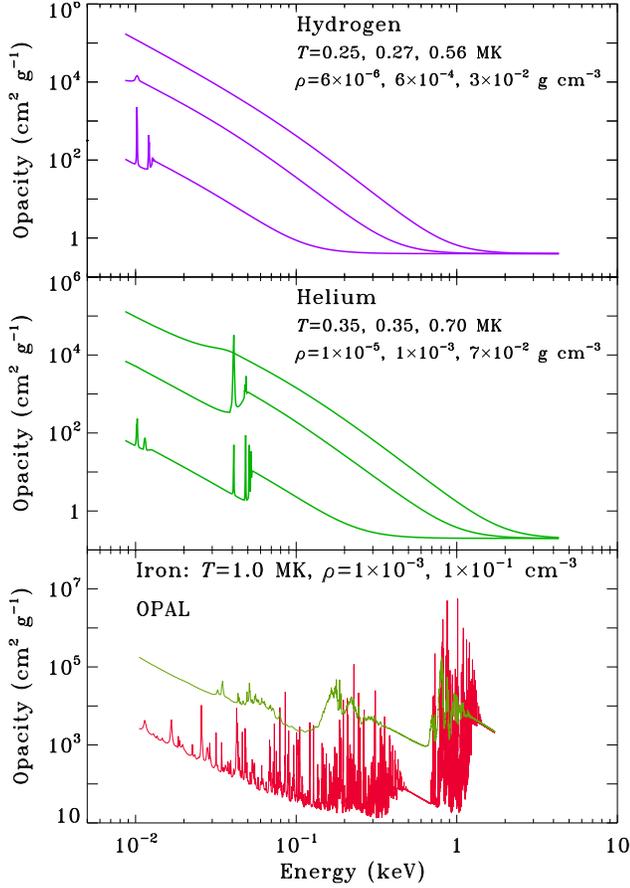,height=12cm,clip=}}
\caption{
Radiative spectral
opacity $k_\nu$ for hydrogen, helium and iron
plasmas at
different temperatures and densities (magnetic field $B=0$). 
The iron opacity is 
provided by the OPAL group
(Rogers \& Iglesias 1994).
}
\label{opac}
\end{figure}

\begin{figure}[!ht]
\centerline{\psfig{file=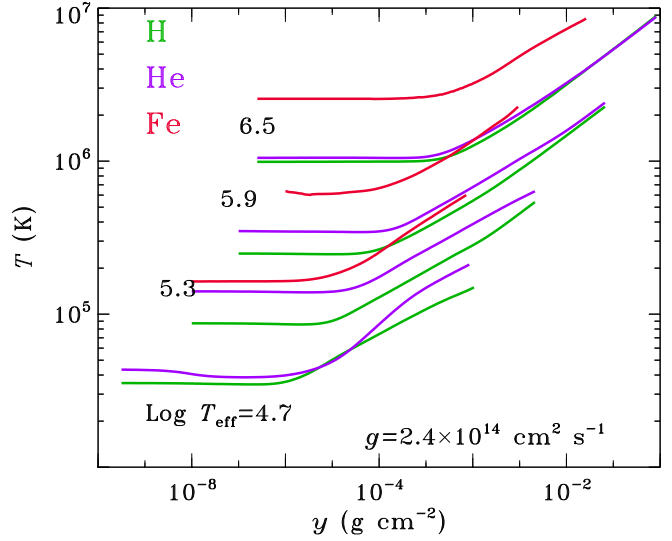,height=8cm,clip=} }
\caption{
Temperature dependences $T(y)$ in nonmagnetic atmosphere
models with different effective temperatures and chemical
compositions.
}
\label{temp}
\end{figure}

The standard approach for the atmosphere modeling includes
solving of three main
equations. The first one is the radiative transfer equation 
for the specific spectral intensity $I_\nu$ (e.~g., Mihalas 1978):
\begin{equation}
\mu\frac{{\rm d}}{{\rm d}y} I_\nu = k_\nu (I_\nu - S_\nu)\, ,
\end{equation}
where $\mu$ is cosine of the angle $\theta$ between the normal to
the atmosphere and the wave-vector of radiation,
$y$ is the column density (${\rm d}y=\rho\, {\rm d}z$, with
$z$ being the geometrical depth),
$k_\nu=\alpha_\nu+\sigma_\nu$ is
the total radiative opacity which includes
the `true absorption'
($\alpha_\nu$) and scattering ($\sigma_\nu$) opacities,
$S_\nu=(\sigma_\nu J_\nu+\alpha_\nu B_\nu)k_\nu^{-1}$ is the source 
function,
$J_\nu=\frac{1}{2}\int_{-1}^1 I_\nu {\rm d}\mu$ is the mean spectral
intensity, and $B_\nu$ is the Planck function.
The boundary condition for this equation is
$I_\nu=0$ for $\mu<0$ at $y=0$,
assuming no incident radiation 
at the NS surface.

NS atmospheres are usually considered to be in radiative
and hydrostatic equilibrium. The first condition implies that
the total energy flux through the atmosphere is constant
and transferred solely by radiation,
\begin{equation}
\int_0^\infty {\rm d}\nu\int_{-1}^1\, \mu I_\nu\, {\rm d}\mu 
=\sigma_{\rm SB}\, \tef^4\, , 
\end{equation}
where $\tef$ is the effective temperature,
and $\sigma_{\rm SB}$ is the Stefan-Boltzmann constant.
The second condition means that the atmospheric pressure is $P=g\, y$
(the radiative force is insignificant unless $\tef\ga 10^7$ K).
Finally, these three equations are supplemented with the equation
of state for the atmospheric plasma and equations of ionization
equilibrium. The latter is needed for computing the electron number density
and the fractions of ions in different stages of ionization
to obtain the radiative opacity with account
for free-free, bound-free and bound-bound
transitions (see examples in Fig.~\ref{opac})\footnote{
Opacities of heavy chemical elements used in this
work were 
calculated by the OPAL group: \\
{\tt http://www-phys.llnl.gov/Research/OPAL/index.html}}.
A detailed description of the model computations can be found, e.g.,
in Zavlin
et al.\ (1996).

\begin{figure*}[ht]
\centerline{\psfig{file=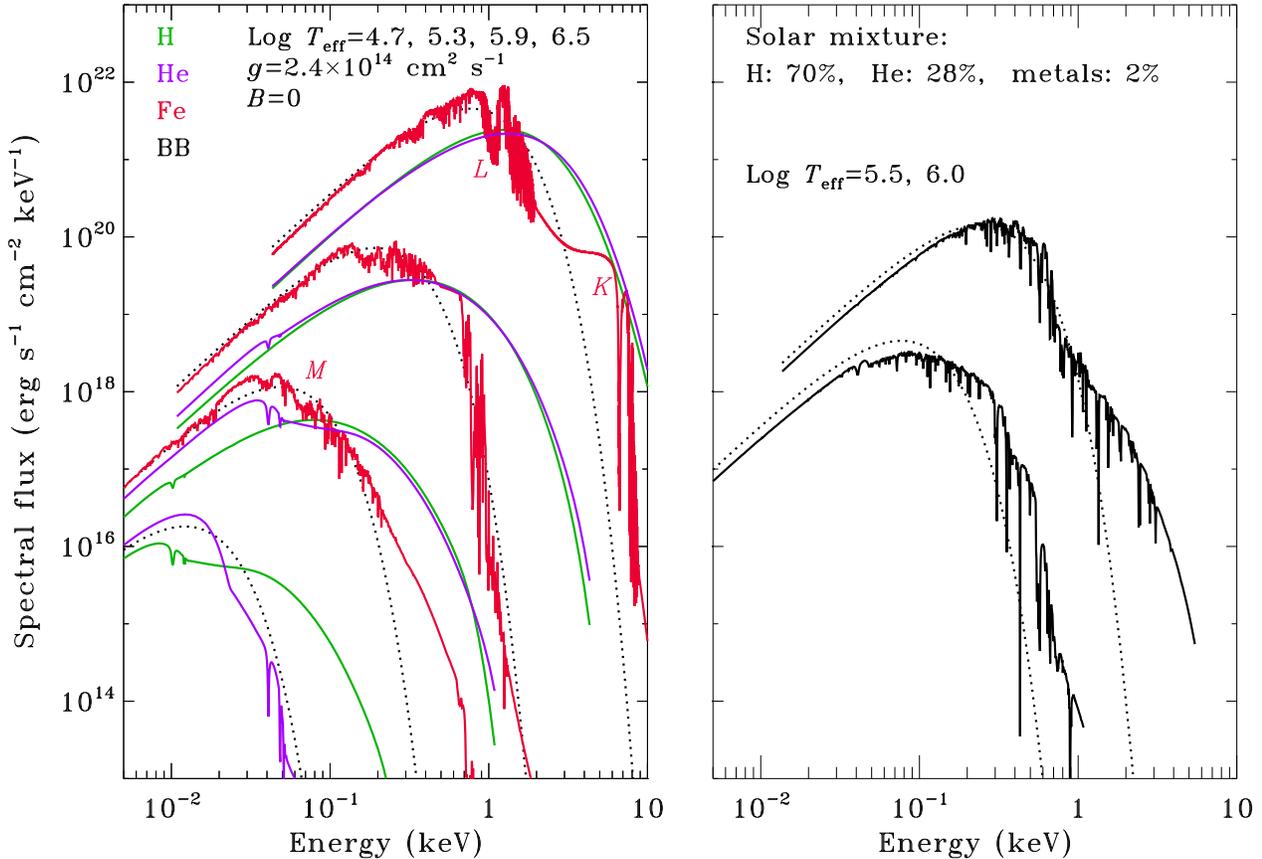,height=12cm,clip=}}
\caption{Spectral fluxes of emergent radiation in nonmagnetic
atmosphere models for several
effective temperatures and chemical compositions.
}
\label{nonmagspec}
\end{figure*}

\subsection{Results}
To explain the main properties of the atmosphere emission, we 
first
present in Figure~\ref{temp}  the most important property
of the atmospheric structure 
responsible for the emergent radiation ---
the depth dependence of the temperature in the NS surface
layers, $T(y)$.

\begin{figure}
\centerline{\psfig{file=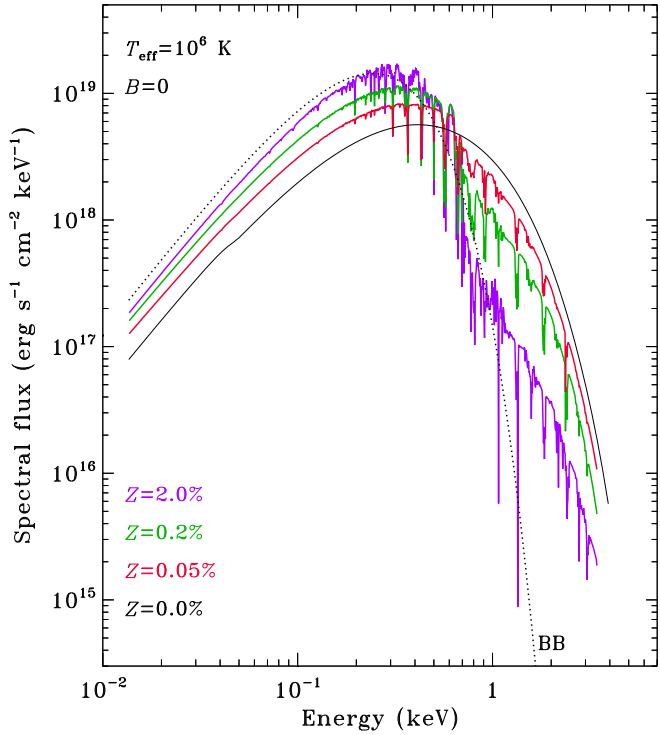,height=10cm,clip=} }
\caption{Spectral fluxes of emergent radiation
in nonmagnetic models with different metal abundances $Z$
($Z=2.0\%$ corresponds to the standard solar mixture
of elements --- see Grevesse \& Noels 1993).
}
\label{zspec}
\end{figure}

\begin{figure}
\centerline{\psfig{file=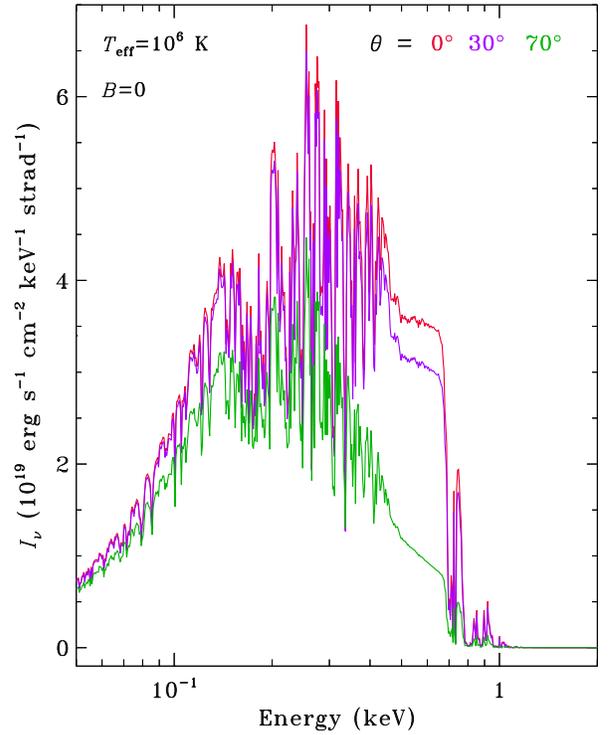,height=10cm,clip=} }
\caption{Spectra of 
specific intensities $I_\nu$
emitted from an iron atmosphere 
with $\tef=1\times 10^6$ K,  at different angles
$\theta$ ($\mu=\cos\theta$). 
}
\label{nonmint}
\end{figure}

Figure~\ref{nonmagspec} presents the spectral fluxes of emergent radiation
at a local surface point,
$F_\nu=\int_0^1 \mu I_\nu {\rm d}\mu$ (at $y=0$),
for several effective temperatures and chemical compositions
(pure hydrogen, helium and iron, and a solar mixture), together
with the blackbody spectral
fluxes 
at the same $\tef$.
The model spectra differ substantially from the blackbody spectra,
particularly in high-energy tails of the spectra of light-element
(hydrogen and helium) atmospheres\footnote{
The nonmagnetic hydrogen atmosphere models are available at
{\tt http://legacy.gsfc.nasa.gov/docs/xanadu/xspec}}. 
The reason for such
behavior is in the rapid decrease of the light-element
opacities with energy ($\sim E^{-3}$, 
see the upper and middle panels in Fig.~\ref{opac}), so that the high-energy
radiation is formed in deeper and hotter layers (Fig.~\ref{temp}).
Since the radiative opacity is large at lower energies,
the low-energy photons escape from very superficial layers with
a temperature  $T(0)<\tef$. Therefore, the low-energy
tails of the model spectra are 
suppressed with respect to
the Rayleigh-Jeans spectra
at the same $\tef$.

The spectra emitted from the heavy-element atmospheres (iron and 
solar mixture) 
exhibit numerous spectral
lines and photoionization edges (e.~g., M, L and K
spectral complexes in the iron spectra)
produced by ions in various ionization stages.
Generally, they are closer to the blackbody
radiation because the energy dependence
of the heavy-element opacities 
is, on average, 
flatter than that for the light elements
(see examples
in the bottom panel of Fig.~\ref{opac}).
However, local deviations from the blackbody spectra
can be very substantial. Moreover, the spectral features
can be rather strong even if metals contribute as little as 0.05\% of
the total NS atmosphere composition (see Fig.~\ref{zspec}).

Although the opacity of the
atmospheric plasma is isotropic in the nonmagnetic case,
the emitted radiation show substantial anisotropy,
i.~e., the specific intensity $I_\nu$ depends on the direction
of emission due to the limb-darkening effect (Fig.~\ref{nonmint}).
At large values of the angle $\theta$ 
between the normal to the surface and the wave-vector
the emerging photons are produced in shallow layers with lower
temperatures.
The anisotropy depends on photon energy and chemical composition
of the atmosphere. This 
effect should be taken 
into account 
to model thermal radiation from a nonuniform
NS surface (see Section~5).

The emergent radiation depends also on the surface gravity:
a stronger gravitational acceleration
increases the density of the atmospheric plasma
and enhances the nonideality effects, 
which results in 
weaker (more smoothed) 
spectral features
(see examples in Fig.~6 of Zavlin et al.\
1996).
However, this effect is rather small and might be important
only for analyzing observational data of extremely good statistics.

\section{Atmosphere models with strong magnetic fields}

It is commonly believed that most NSs (at least active radio/X-ray/$\gamma$-ray
pulsars) have rather strong surface magnetic fields,
$B\sim 10^{10}$--$10^{14}$ G.
Such fields drastically change the properties of NS atmospheres,
as discussed above.
Magnetic hydrogen models have been 
developed
by Shibanov et al.~(1992), Pavlov et al.~(1994), Shibanov \& Zavlin (1995),
Pavlov et al.~(1995), and Zavlin et al.~(1995b).
These models use
simplified radiative opacities of strongly magnetized,
partially ionized plasma, which
do not include the
bound-bound transitions.
However, they
are considered to be
reliable enough in the case of high temperatures
($\tef\ga 
10^6$ K at typical radio-pulsar fields 
$\sim 10^{12}$ G), when the atmospheric
plasma is 
almost fully ionized even in the strong
magnetic fields.
Recently, 
completely ionized hydrogen models for superstrong magnetic fields,
$B\sim 10^{14}$--$10^{15}$ G, have been 
presented in a number of papers
(Bezchastnov et al.\ 1996; Bulik \& Miller 1997;
 \"Ozel 2001; Ho \& Lai 2001; 
Zane et al.\ 2001), concerned mainly with the vacuum polarization
effects (Pavlov \& Gnedin 1984) and the ion cyclotron
lines whose energies get into the X-ray band at $B\ga 2\times 10^{13}$ G.
A set of magnetic iron models was constructed by Rajagopal,
Romani \& Miller (1997), with the use of rather crude
approximations for the very complicated properties of iron ions in
strong magnetic fields. The iron atmosphere
spectra show many
prominent spectral features which could be very useful
to measure the NS magnetic field and the mass-to-radius
ratio.

\subsection{General approach}

\begin{figure*}[ht]
\centerline{\psfig{file=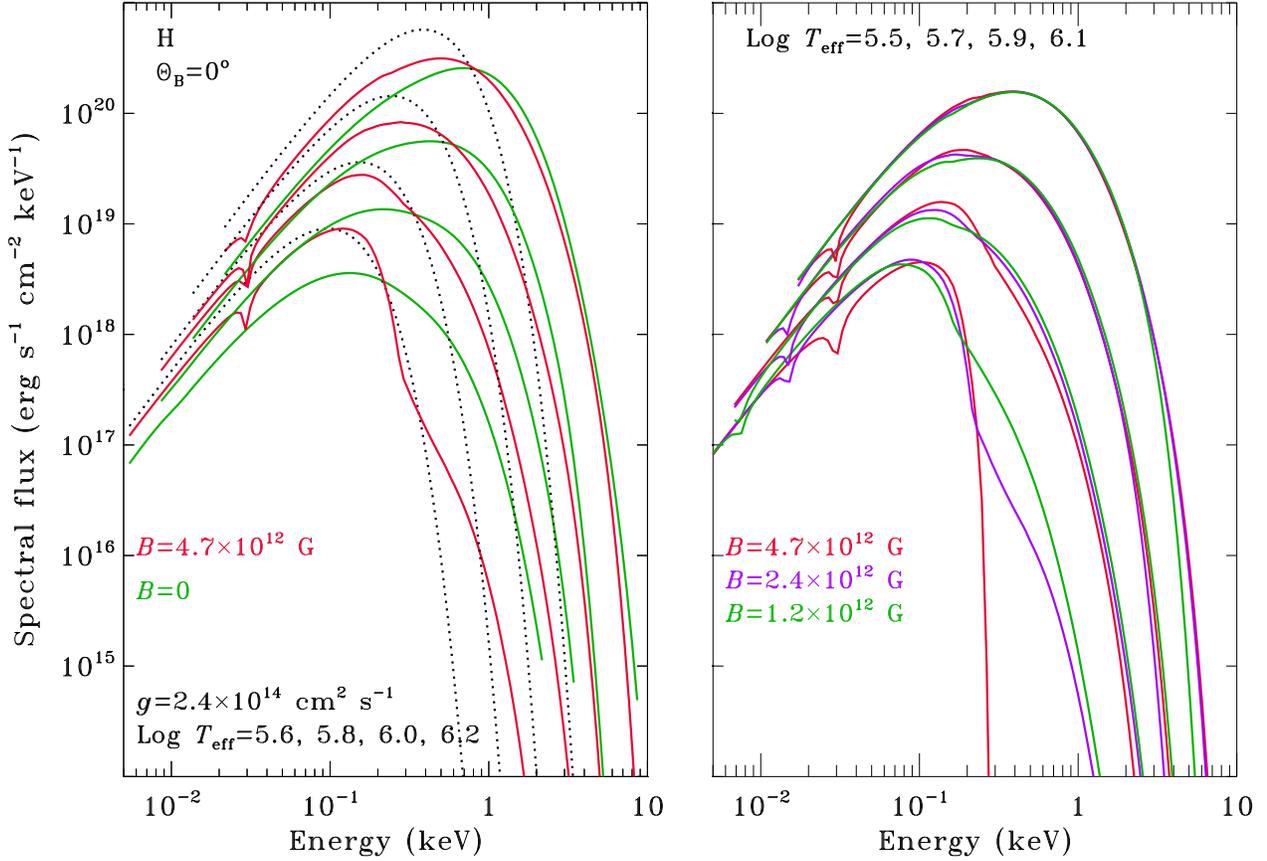,height=12cm,clip=}}
\caption{Spectral fluxes of 
radiation emerging from magnetic hydrogen
atmospheres,
for several
effective temperatures and magnetic fields perpendicular to
the surface.
}
\label{magspec}
\end{figure*}

Modeling magnetic NS atmospheres is generally 
similar
to the nonmagnetic case. The main difference
is that the atmospheric radiation is polarized, and the
radiative opacities depend on the polarization and direction of 
radiation. Gnedin \& Pavlov (1974) described the radiative transfer
in a strongly magnetized plasma in terms of coupled
equations for specific intensities of
two normal modes, 
$I_{\nu,1}$ and
$I_{\nu,2}$, with different polarizations and
opacities: 

$$
\mu\frac{{\rm d}}{{\rm d}y} I_{\nu,j}(\vec{n}) =
k_{\nu,j}(\vec{n}) I_{\nu,j}(\vec{n}) - 
$$
$$
- \left[\sum_{i=1}^2\oint\, {\rm d}\vec{n'}\,  
I_{\nu,i}(\vec{n'})\, \sigma_{\nu,ij}(\vec{n'},\vec{n})
+ \alpha_{\nu,j}(\vec{n})\frac{B_\nu}{2}\right],~~~~~~(3)
$$ 
where $\vec{n}$ is the (unit) wave-vector,
$\alpha_{\nu,j}$ is the absorption opacity for the $j$-th
mode, and $\sigma_{\nu,ij}$ is the scattering opacity
from mode $i$ to mode $j$. It should be noted that the opacity
depends 
on the angle between the wave-vector and
the magnetic field, so that $I_\nu$ depends not only on $\theta$,
but also on $\Theta_{\rm B}$, the angle between
the local magnetic field and the normal to the NS surface.
Similar to the nonmagnetic case, equations~(3) are
supplemented with the equations of radiative and
hydrostatic equilibrium. To overcome the problems
with the sharp angular dependence
of the radiative opacities (Kaminker, Pavlov \& Shibanov 1982),
a two-step method for 
modeling of magnetic NS atmospheres was developed (Pavlov et al.~1994;
Shibanov \& Zavlin 1995). At the first step, the radiative transfer
is solved in the diffusion approximation for the mean
intensities $J_{\nu,j}=(4\pi)^{-1}\, \oint I_{\nu,j}(\vec{n})\, {\rm d}\vec{n}$: 
\setcounter{equation}{3}
\begin{equation}
\frac{{\rm d}}{{\rm d}y} d_{\nu,j} \frac{{\rm d}}{{\rm d}y} J_{\nu,j}=
\bar{\alpha}_{\nu,j}\left[J_{\nu,j}-\frac{B_\nu}{2}\right]+
\bar{\sigma}_\nu\left[J_{\nu,j}-J_{\nu,3-j}\right]\, ,
\end{equation}
where $d_{\nu,j}$ is the diffusion coefficient,
$\bar{\alpha}_{\nu,j}$ and 
$\bar{\sigma}_\nu$ are the angle-averaged absorption
and scattering opacities
(see Pavlov et al.~1995 for details).
Next, the atmospheric structure obtained at the first step
is corrected using an iterative procedure applied to the
exact equations of the radiative transfer.

\subsection{Results}

\begin{figure}
\centerline{\psfig{file=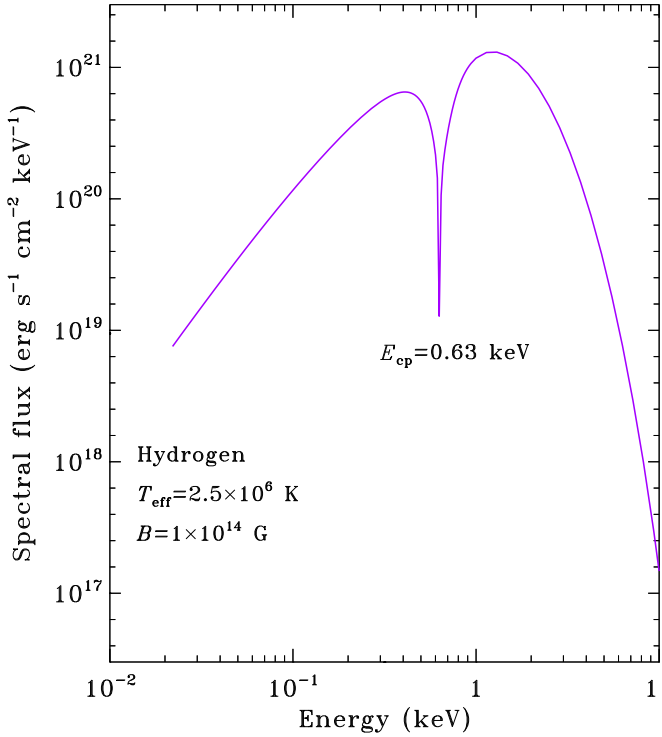,height=10cm,clip=} }
\caption{Spectral flux of 
radiation
emerging from an atmosphere with $B=10^{14}$ G, with
a strong proton cyclotron line.
}
\label{protline}
\end{figure}

\begin{figure}
\centerline{\psfig{file=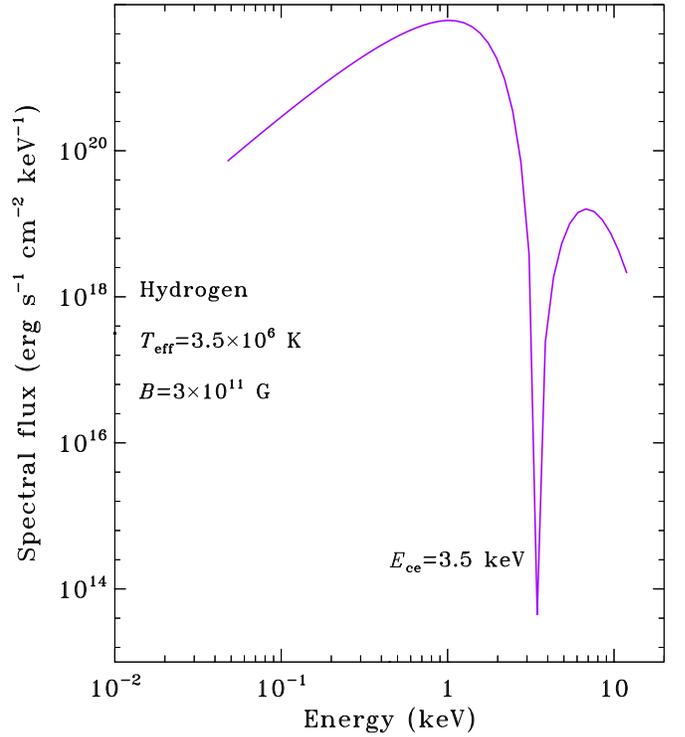,height=10cm,clip=} }
\caption{Spectral flux of 
radiation emerging from an
atmosphere 
with $B=3\times 10^{11}$ G, with
a strong
electron cyclotron line.
}
\label{elecline}
\end{figure}

Figure~\ref{magspec} shows polarization-summed spectral
fluxes of the emergent radiation, 
$F_\nu=\int_0^1\, \mu (I_{\mu,1}+I_{\mu,2})\, {\rm d}\mu$,
emitted by a local element of the NS surface,
for several values of effective temperature and magnetic field
perpendicular to the surface 
($\Theta_{\rm B}=0$).
The main result is that the magnetic model spectra
are 
harder than the blackbody radiation of
the same $\tef$, although they are softer than the low-field
spectra.
Similar to the low-field
case, this is explained
by the temperature growth with depth and
the opacity decrease
at higher energies, which is more gradual
($\propto E^{-1}$ for the mode
with smaller opacity)
compared to the nonmagnetic case. 
At lower effective temperatures, $\tef\la 10^6$ K,
the photoionization opacity is important, and the dependence
of the spectra on magnetic field becomes more pronounced
(see the right panel in Fig.~\ref{magspec}).
The field dependence is 
most clearly seen, at $\tef$ sufficiently high to provide
strong ionization, in the
proton cyclotron lines centered at energies 
$E_{\rm cp}=6.3\, (B/10^{12}\, {\rm G})$ eV.
If the magnetic field is very large, $B\ga 10^{14}$ G,
the proton cyclotron line shifts into the X-ray band and
can be very strong (see Fig.~\ref{protline}). On the other hand, if 
the magnetic field is not so large, $B=10^{10}$--$10^{12}$ G,
the NS atmosphere spectra may exhibit
the electron cyclotron lines
in the X-ray band,
at $E_{\rm ce}=11.6\, (B/10^{12}\, {\rm G})$ keV (see Fig.~\ref{elecline}).

\begin{figure}
\centerline{\psfig{file=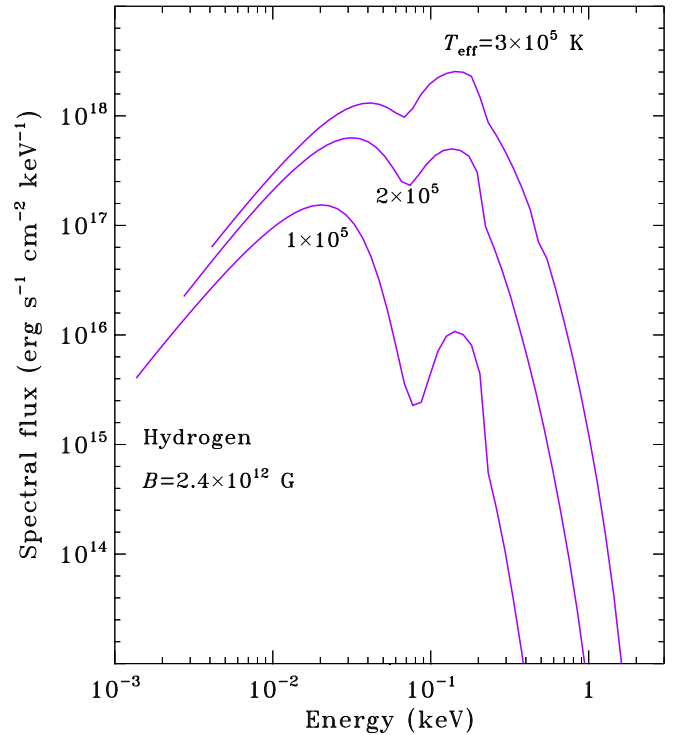,height=10cm,clip=} }
\caption{Spectral fluxes of 
radiation emerging from
hydrogen atmospheres
with $B=2.4\times 10^{12}$ G and
different values of $\tef$, with 
bound-bound transitions included.
}
\label{bbspec}
\end{figure}

\begin{figure}
\centerline{\psfig{file=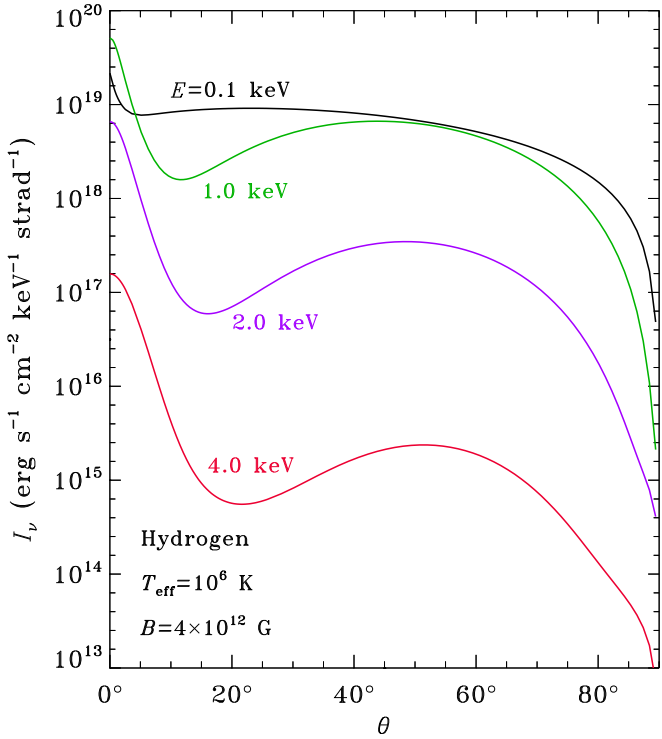,height=10cm,clip=} }
\caption{Angular dependences of polarization-summed specific intensities
at different photon energies $E$ in the hydrogen
atmosphere model with $\tef=1\times 10^6$ K and magnetic field
$B=4\times 10^{12}$ G, perpendicular to the surface.
\label{magint}
}
\end{figure}

First calculations of hydrogen atmosphere
models which include bound-bound transitions 
show that spectral lines, considerably broadened by
the motional Stark effect (Pavlov \& M\'esz\'aros 1993; Pavlov
\& Potekhin 1995)
become prominent at $\tef\la 5\times 10^5$ K. The strongest line
corresponds to the transition between the ground state and
the lowest excited state; its energy is
$E\approx [75 (1+0.13 \ln(B/10^{13}\, {\rm G}) + 
63(B/10^{13}\, {\rm G})]$ eV,
at $B\sim 10^{13}$ G (see examples in Fig.~\ref{bbspec}).

Radiation emerging
from a magnetized NS atmosphere is
strongly anisotropic. Angular dependences of
the local specific intensities, $I_\nu=I_{\nu,1}+I_{\nu,2}$,
show a complicated ``pencil-plus-fan'' structure ---
a narrow peak along the direction of the magnetic field
(where the atmospheric plasma is most transparent for
the radiation),
and a broader peak at intermediate angles. The widths
and strengths of the peaks depend on magnetic field and photon energy 
(see examples in Fig.~\ref{magint}). Obviously, it is very important
to account for this anisotropy while modeling the
radiation from 
a NS
with nonuniform surface
magnetic field, effective temperature, and chemical composition.

\section{Thermal radiation as seen
by a distant observer}

Results presented in Sections~3 and 4 describe spectral 
radiation emitted by a {\sl local} element at the NS surface.
The effective temperature and/or magnetic 
field distributions over the NS surface can be nonuniform
(for example, if a NS has a dipole magnetic filed,
the 
effective temperature 
decreases from the magnetic poles to the equator). 
To calculate
the {\sl total} NS emission, one has to integrate
the local intensities, computed for 
local temperatures
and magnetic fields, over the visible part of the NS 
surface $S$, with account for the 
gravitational
redshift and bending of photon trajectories:
\begin{equation}
F(E) = g_{\rm r}\frac{1}{D^2}\int_S \mu\, 
I(E/g_{\rm r})\, {\rm d}S\, , 
\end{equation}
where $D$ is the distance from the NS to the observer,
and $E$
is the {\em observed} (redshifted) 
energy\footnote{
To take into account the interstellar absorption,
a factor of $\exp[-n_{\rm H} \sigma_{\rm eff}(E)]$
should be added in eq.\ (5)
[$\sigma_{\rm eff}(E)$ is the absorption cross section per
hydrogen atom].
}.
More details about the integration over the 
NS surface  can be found in Pavlov \& Zavlin (2000b).
In should be noted that if a NS has a nonuniform (e.g., dipole) distribution
of the magnetic field,
the integration
broadens the spectral features (in particular, cyclotron
resonance lines --- see Zavlin et al.~1995b;
Pavlov \& Zavlin 2000b; Zane et al.\ 2001).

\begin{figure}
\centerline{\psfig{file=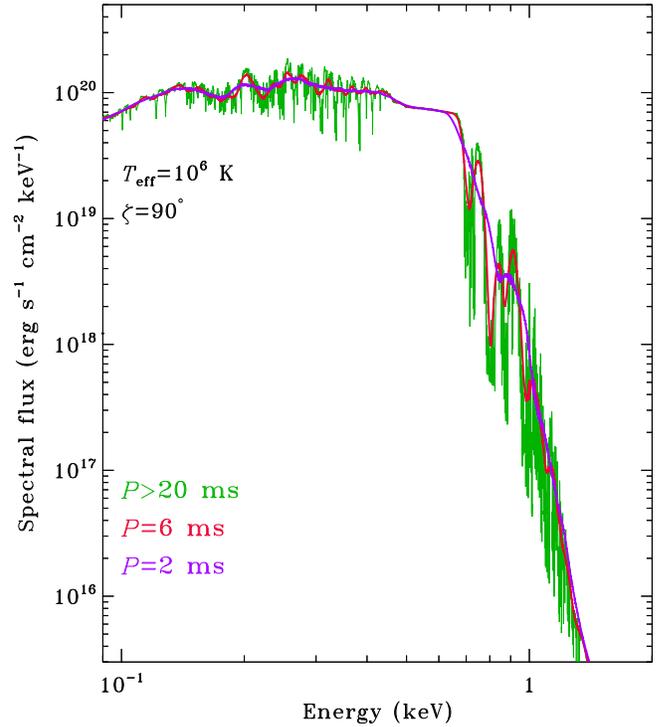,height=10cm,clip=} }
\caption{
Spectral fluxes of 
radiation from the {\sl whole} surface
of a nonmagnetic NS covered with an iron atmosphere at $\tef=1\times 10^6$ K.
The (unredshifted) spectra are 
calculated for different NS rotation periods $P$,
at the inclination of rotational axis
$\zeta=90^\circ$.
}
\label{fesmooth}
\end{figure}

\begin{figure*}[ht]
\centerline{\psfig{file=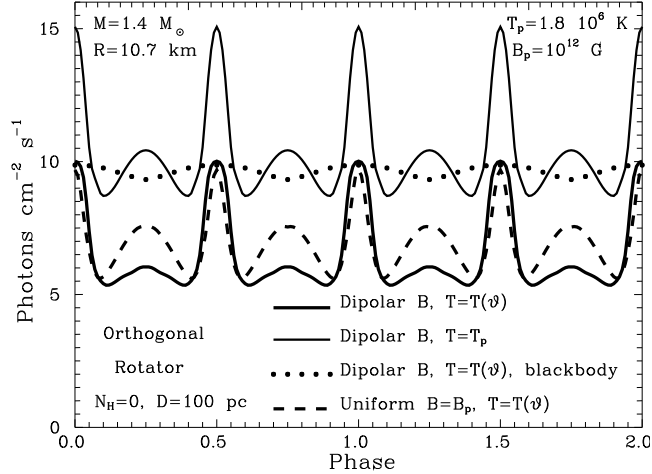,height=7cm,clip=} }
\caption{
Energy-integrated light curves of a rotating NS 
with different distributions of surface temperature and magnetic field 
(from Shibanov et al.\ 1995).
The temperature is either uniform, $T=T_{\rm p}$, or
nonuniform, $T=T(\vartheta)$,
as given by computations with account for the anisotropy
of the heat conductivity in the NS crust
($\vartheta$ is magnetic colatitude,
$T_{\rm p}$ and 
$B_{\rm p}$ are the effective temperature and the field at the
magnetic poles).
The NS rotational axis is perpendicular to both
the magnetic axis and the line of sight ($\alpha=\zeta=90^\circ$).
\label{maglc}
}
\end{figure*}

If the NS is a fast rotator, one should take into account
the Doppler shifts of energies of
photons emitted from surface elements moving with
different radial velocities.
Maximum values of these velocities, 
$v_r=2\pi R P^{-1} \sin\zeta$ ($P$ is the rotation period,
$\zeta$ the inclination of the rotation axis),
can be as high as 10\%--15\% of the speed of light for millisecond
periods.
Figure~\ref{fesmooth} presents (unredshifted)
spectra emitted from the whole NS surface
covered with an iron atmosphere (nonmagnetic case). 
For slowly rotating NSs, $P\ga 20$ ms, there is no significant
effect 
on the emergent radiation. Fast rotation, $P\la 10$ ms,
may lead to complete smearing of 
weak and narrow spectral lines, provided $\sin\zeta$ is large enough,
leaving only most prominent spectral jumps around the strongest
photoionization edges.

In the case of radio pulsars, 
a fraction of the observed thermal X-ray radiation is emitted from
small hot spots (polar caps)
around the NS magnetic poles.
These polar caps
can be heated up to X-ray temperatures by relativistic
particles impinging on the 
poles from
the acceleration zone in the pulsars' magnetosphere. 
In the case of millisecond pulsars, with characteristic ages of
$10^8$--$10^9$ yr, the whole NS surface (except for the polar caps)
is believed to be 
too cold ($\tef\la 10^5$ K) to be observable in X-rays,
so the detected thermal emission can be modeled as 
emitted from small heated spots:
\begin{equation}
F(E)=g_{\rm r}\frac{S_{\rm a}}{D^2}
I(E/g_{\rm r},\, \theta^*)\, ,
\end{equation}
where the apparent spot area $S_{\rm a}$ and the angle
$\theta^*$ between the wave-vector 
and the radius-vector to the hot spot are computed
with account for the effect of gravitational 
bending.
These quantities depend on
the angles $\alpha$
(between the rotational and magnetic axes)
 and $\zeta$ (between the rotational axis and the line of sight
--- see Fig.~\ref{sketch}), and
the gravitational parameter $g_{\rm r}$
(see Zavlin et al.\
1995a for details).

The flux given by equations (5) or (6)
varies with the period of NS rotation.
One can obtain a large variety of 
pulse profiles 
at different assumptions on the 
angles $\alpha$ and $\zeta$ 
and 
the NS mass-to-radius ratio.
Examples of pulse profiles computed for radiation
from the whole NS surface are shown in Figure~\ref{maglc}; 
pulse profiles of thermal radiation from heated polar
caps are presented by Zavlin et al.\ (1995a).

Although the model atmosphere spectra are different
from the blackbody spectra, very often an observed thermal spectrum
can be fitted equally well with the blackbody and NS atmosphere
models, particularly when the energy resolution is low and/or
the energy band is narrow.
However, the parameters obtained from such fits are quite different, 
especially when the hydrogen or helium atmospheres are used.
Since the light-element atmosphere spectra are much harder
than the blackbody spectra
at the same effective temperature,
atmosphere model fits
give temperatures $T_{\rm atm}$ significantly lower
than 
the blackbody temperature $T_{\rm bb}$,
with a typical ratio $T_{\rm bb}/T_{\rm atm}\sim 2$--3
(see, for example, Pavlov et al.~1996; Pons et al.~2002).
On the other hand, to provide the same total energy flux,
the blackbody fit gives a smaller 
normalization factor, proportional to $S/D^2$,
than the  atmosphere model fit.
In other words, the light-element atmosphere
fit gives a considerably larger size
of the emitting region, $S_{\rm atm}/S_{\rm bb}\sim 50$--200,
for the same distance to the source.

\section{Conclusions}

In the recent decade,
substantial progress has been made in modeling atmospheres of
isolated NSs. Best investigated cases are nonmagnetic
atmospheres and fully-ionized light-element atmospheres
with strong magnetic fields.
The atmosphere models
have been applied to the interpretation of thermal emission from NSs of
different types.
For instance, Pavlov \& Zavlin (1997), Zavlin \& Pavlov (1998), and
Zavlin et al.\ (2002) analyzed the
X-ray emission
from the millisecond pulsar J0437--4715 using the nonmagnetic light-element
atmosphere models;
Pavlov et al.\ (2001) applied the magnetic hydrogen models
to the analysis
of radiation from
the Vela pulsar; 
Zavlin, Pavlov \& Tr\"umper (1998)
and Zavlin, Tr\"umper \& Pavlov (1999) 
used the magnetic light-element
models
to interpret the thermal emission from NSs in the supernova
remnants PKS 1209--51/52 and Puppis~A.
More examples are presented in the contribution by Pavlov et al.\
in this volume.

However, 
a number of problems in the atmosphere
modeling remains to be solved.
First of all,
investigations of the structure of various atoms, molecules,
and molecular chains in strong magnetic fields, 
as well as radiative transitions in these species (Pavlov 1998),
are necessary
to construct magnetic atmosphere models of different chemical compositions.
First efforts in this direction are being undertaken (Mori \& Hailey 2002).
Particularly interesting are the
(virtually unknown) radiative properties of matter in
superstrong magnetic fields, $B\ga 10^{14}$ G,
apparently found in anomalous X-ray pulsars and soft gamma-ray repeaters.
Further work is needed on radiative properties
of nonideal plasmas 
and condensed matter. 
These investigations will eventually result in more
advanced models for thermal radiation of isolated NS, needed
for the interpretation
of high-quality
observations of these objects with the 
{\sl Chandra} and {\sl XMM}-Newton X-ray observatories.

\begin{acknowledgements}
The authors gratefully acknowledge the support by the Heraeus foundation
and the hospitality of the Physikzentrum Bad Honnef.
Work of GGP was partially supported by NASA grant NAG5-10865.
\end{acknowledgements}

\end{document}